\documentclass{article}
\usepackage{spconf,amsmath,graphicx,hyperref}
\usepackage{comment}
\usepackage{bm}
\usepackage{amsmath}
\usepackage{amssymb}
\usepackage{amsfonts}
\usepackage{xcolor}
\usepackage{multirow}

\newcommand{\sedit}[1]{\textcolor{black}{#1}}
\newcommand{\ssedit}[1]{\textcolor{black}{#1}}
\newcommand{\sssedit}[1]{\textcolor{black}{#1}}
\newcommand{\kedit}[1]{\textcolor{black}{#1}}
\newcommand{\redit}[1]{\textcolor{black}{#1}}
\newcommand{\nsedit}[1]{\textcolor{black}{#1}}

\title{Wave-Trainer-Fit: Neural Vocoder with Trainable Prior and Fixed-Point Iteration towards High-Quality Speech Generation from SSL features}
\name{Hien Ohnaka$^{1,}$\sthanks{Work done during part-time job in LY Corporation.}, Yuma Shirahata$^2$, Masaya Kawamura$^2$}
\address{
  $^1$Nara Institute of Science and Technology, Nara, Japan\\
  $^2$LY Corporation, Tokyo, Japan
}
\begin{document}
\ninept
\maketitle
\begin{abstract}
We propose WaveTrainerFit, a neural vocoder that performs high-quality waveform generation from data-driven features such as SSL features.
WaveTrainerFit builds upon the WaveFit vocoder, which integrates diffusion model and generative adversarial network. 
\redit{Furthermore, the proposed method incorporates the following key improvements: }
1. By introducing trainable priors, the inference process starts from noise close to the target speech instead of Gaussian noise.
2. Reference-aware gain adjustment is performed by imposing constraints on the trainable prior to matching the speech energy. 
These improvements are expected to reduce the complexity of waveform modeling from data-driven features, enabling high-quality waveform generation with fewer inference steps.
Through experiments, we showed that WaveTrainerFit can generate highly natural waveforms with improved speaker similarity from data-driven features, while requiring fewer iterations than WaveFit.
\ssedit{Moreover, we showed that the proposed method works robustly with respect to the depth at which SSL features are extracted.}
\nsedit{Code and pre-trained models are available from \url{https://github.com/line/WaveTrainerFit}.}
\end{abstract}
\begin{keywords}
Neural vocoder, self-supervised learning, variational autoencoder, speech synthesis.
\end{keywords}
\section{Introduction} \label{sec:intro}
Neural vocoders are deep neural networks (DNNs) that generate speech waveforms from acoustic features.
Recently, they have become indispensable components in various tasks such as text-to-speech~\cite{tacotron,fastspeech2}, voice conversion~\cite{vc_survey}, and speech-to-speech translation~\cite{translacotron2}.
Specifically, neural vocoders from Mel-spectrograms have been widely researched, and they have become capable of generating waveforms of the same quality as human speech~\cite{wavefit,hifigan}.

In recent years, the advancement of self-supervised learning (SSL) models~\cite{wavlm,xls_r,hubert} has changed the way of conventional speech processing tasks.
SSL models have demonstrated high performance in generative applications such as text-to-speech synthesis~\cite{ssl_study_read_sponte_tts,hundred_multilingual_tts}, and speech restoration~\cite{miipher,miipher2}. 
Their advantage is the ability to learn data-driven features that represent speech signals from large amounts of unlabeled data. 
In light of this background, robust vocoders from SSL features are becoming increasingly essential.

WaveFit~\cite{wavefit,miipher} is a fixed-point iteration vocoder that combines generative adversarial networks (GANs)~\cite{hifigan,parallelwavegan,mb_melgan,melgan} with diffusion model~\cite{specgrad,wavegrad}-like inference.
Due to the combination of the two generative models, WaveFit can generate high-quality waveforms \sssedit{from not only Mel-spectrograms, but also SSL features} that are not specialized for waveform generation tasks~\cite{miipher,miipher2,libritts_r,fleurs_r}.

\sssedit{When Mel-spectrograms are used as input of WaveFit~\cite{wavefit}, signal processing knowledge could be applied for effective training of diffusion models in two ways: 1. initial noise could be sampled from a hand-crafted prior distribution using the spectral envelope~\cite{specgrad}, rather than the Gaussian noise. 2. Gain adjustment is done explicitly using the power of the Mel-Spectrogram. }
\sssedit{However, when SSL features are used as input~\cite{miipher}, these methods cannot be used since SSL features are not derived from signal-processing.}

\begin{figure}[t!]
\begin{center}
\includegraphics[width=\linewidth]{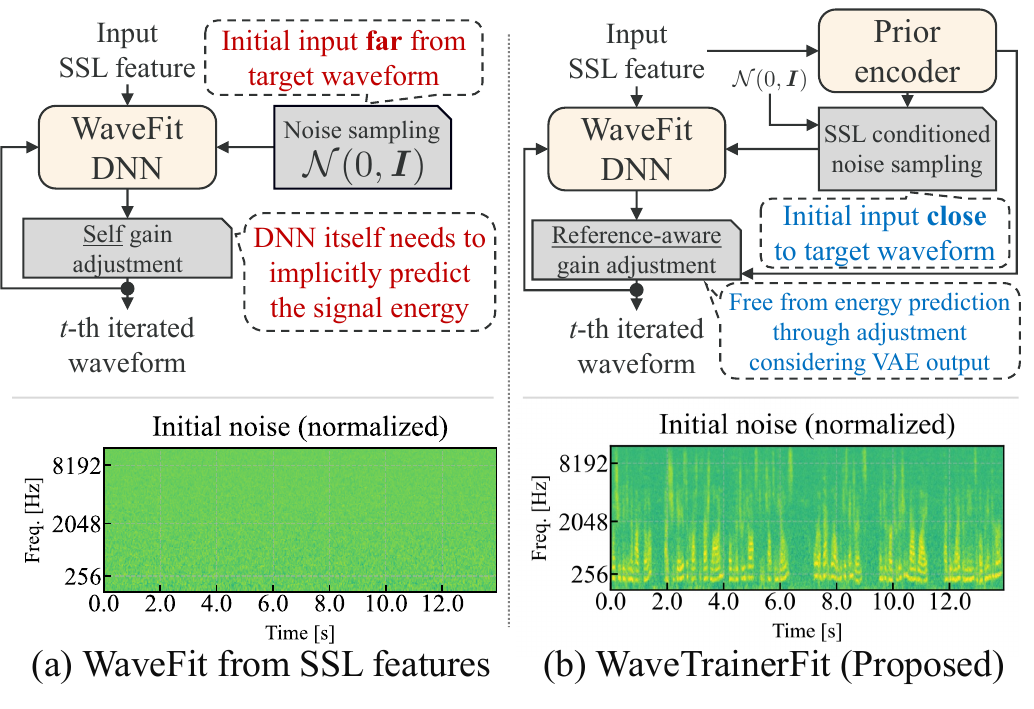}
\vspace{-9.5mm}
\caption{\sedit{
Conceptual diagrams and noise examples of each methods.
The bottom images show the log Mel-spectrograms of initial noise.
}} \label{fig:concept}
\vspace{-9mm}
\end{center}
\end{figure}

To overcome these limitations, we propose a neural vocoder with \textbf{\underline{train}}abl\textbf{\underline{e}} prio\textbf{\underline{r}} and \textbf{\underline{f}}ixed-point \textbf{\underline{it}}eration (\textbf{\underline{WaveTrainerFit}}) for improved waveform generation from SSL features. 
The key improvements in WaveTrainerFit over WaveFit are illustrated in Fig.~\ref{fig:concept}.
First, by introducing variational autoencoder (VAE)~\cite{vae}-based trainable priors, \sedit{we achieve sampling of noise close to target waveform.
Since inference can start from a point close to speech, high-quality waveform generation with fewer iterations could be expected.}  
Furthermore, by imposing constraints on the priors to learn the energy of speech, we realize gain adjustment \sssedit{using the energy of the priors}
, which frees the vocoder from the implicit energy inference task.
As a result, the model can focus on more important aspects of waveform modeling, and is thought to mitigate the difficulty of training.

Objective evaluations showed that WaveTrainerFit achieves better quality \ssedit{waveform} generation compared to WaveFit with fewer iterations.
Moreover, the proposed method can generate natural waveforms even when conditioned on SSL features from deep layers, which contain limited acoustic information.
Finally, subjective evaluation showed that the proposed method outperforms the baselines in speaker similarity.
Audio samples are available on our demo page~\footnote{\tiny \url{https://i17oonaka-h.github.io/projects/research_topics/wave_trainer_fit/}}.

\section{Related work}
\subsection{WaveFit: Neural vocoder with fixed-point iteration} \label{sec:ssl-wavefit}
WaveFit~\cite{wavefit,miipher} is an iterative-style non-autoregressive neural vocoder that combines the iterative processing of diffusion models with GAN-based loss. \sssedit{Although WaveFit has variants that take both Mel-spectrograms and SSL features as inputs, in this work we focus on the latter.}
This method generates the speech waveform $\bm{y}_0 \in \mathbb{R}^D$ from \sssedit{Gaussian noise $\bm{y}_{T} \in \mathbb{R}^D \sim \mathcal{N}(0,\bm{I})$} and SSL features 
$\bm{c}$ by applying $T$ denoising mapping processes~\cite{miipher}:
\begin{align}
    \bm{y}_{t-1}&=\hat{\mathcal{G}}(\bm{z}_t),\quad \bm{z}_t=\bm{y}_t - \mathcal{F}_\theta(\bm{y}_t,\bm{c},t), \\
    \hat{\mathcal{G}}(\bm{z}_t)&=\beta_\mathrm{scale} \cdot \bm{z}_t/\max(\mathrm{abs}(\bm{z}_t)), \label{eq:self_gain_adjustment}
\end{align}
where $D$ is the number of time-domain samples, $\mathcal{F}_\theta$ is a DNN trained to estimate noise components, 
$\beta_\mathrm{scale}$ is a scaling factor, and $\mathrm{abs}(\cdot)$
returns element-wise absolute values of the input vector.
$\hat{\mathcal{G}}(\bm{z}_t)$ is a self-gain adjustment operator that normalizes the power of predicted waveform.
The loss function $\mathcal{L}^\mathrm{WF}$ is represented by the following equation:
\begin{align}
    \mathcal{L}^\mathrm{WF}\!\!&=\!\frac{1}{T}\!\sum_{t=0}^{T-1}\mathcal{L}_\mathrm{G}^\mathrm{gan}(\bm{x}_0,\bm{y}_t)\!+\!\mathcal{L}_\mathrm{D}^\mathrm{gan}(\bm{x}_0,\bm{y}_t)\!+\!\lambda_\mathrm{S}\mathcal{L}^\mathrm{S}(\bm{x}_0,\bm{y}_t). \label{eq:wavefit_loss}
\end{align}
Here, $\bm{x}_0$ is a target waveform, 
\sssedit{$\mathcal{L}_\mathrm{G}^\mathrm{gan}$ and $\mathcal{L}_\mathrm{D}^\mathrm{gan}$ are the loss functions of the generators and discriminators in~\cite{wavefit},}
$\lambda_\mathrm{S}$ is the weight parameter, and $\mathcal{L}^\mathrm{S}$ is the multi-resolution STFT loss~\cite{parallelwavegan,mb_melgan}.

WaveFit has been successfully employed as a vocoder from SSL features in tasks such as speech restoration~\cite{miipher,miipher2} and high-quality dataset creation~\cite{libritts_r,fleurs_r}. However, since WaveFit was originally designed as a vocoder for the Mel-spectrogram, a signal-processing-based acoustic feature, it makes the following two compromises when using data-driven SSL features as input:
\\\textbf{1. Noise sampling.} In the initial noise sampling of the diffusion model, it cannot use hand-crafted noise sampling methods~\cite{specgrad,priorgrad}, which leverages signal-processing knowledge derived from the Mel-spectrogram. Therefore, it resorts to simple Gaussian noise.
\\\textbf{2. Gain adjustment.} In gain adjustment, it cannot utilize the power obtained from the Mel-spectrogram. Therefore, it must instead be learned by the model as in (\ref{eq:self_gain_adjustment}).

\subsection{RestoreGrad: Diffusion model with trainable prior}
\label{sec:restoregrad}
RestoreGrad~\cite{restoregrad} is a method to obtain an informative prior for diffusion models even when hand-crafted noise sampling methods are difficult.
To achieve this, RestoreGrad combines diffusion models with trainable priors, which are modeled by VAEs. 
\sssedit{Specifically, RestoreGrad is trained to minimize the \kedit{Kullback-Leibler (KL)} divergence between the posterior distribution $\mathcal{N}(0, \bm{\Sigma}_\mathrm{post})$ and the prior distribution $\mathcal{N}(0, \bm{\Sigma}_\mathrm{prior})$}:
\begin{align}
    \mathcal{L}^\mathrm{PM}(\bm{\Sigma}_\mathrm{post},\bm{\Sigma}_\mathrm{prior})&=\log \frac{|\bm{\Sigma}_\mathrm{prior}|}{|\bm{\Sigma}_\mathrm{post}|} + \mathrm{tr}(\bm{\Sigma}_\mathrm{prior}^{-1}\bm{\Sigma}_\mathrm{post}) \label{eq:L_PM_restoregrad}.
\end{align}
$\bm{\Sigma}_\mathrm{prior}$ and $\bm{\Sigma}_\mathrm{post}$ are the covariance matrix generated by the prior encoder $\mathcal{V}_\mathrm{prior}(\bm{c})$ and the posterior encoder $\mathcal{V}_\mathrm{post}(\bm{c},\bm{x}_0)$, respectively. $\bm{x}_0$ is the target waveform, and $\bm{c}$ is the conditional feature.
\sssedit{To help the posterior encoder learn the informative representation, additional loss term $\mathcal{L}^\mathrm{LR}(\bm{x}_0,\bm{\Sigma}_\mathrm{post})$ is introduced:}

\begin{align}
    \mathcal{L}^\mathrm{LR}(\bm{x}_0,\bm{\Sigma}_\mathrm{post}) &= \log{|\bm{\Sigma}_\mathrm{post}|} + \overline{\alpha}_T \bm{x}_0^T \bm{\Sigma}_\mathrm{post}^{-1}\bm{x}_0, \label{eq:L_LR_restoregrad}
\end{align}
where  $\overline{\alpha}_T$ is a weight based on the variance schedule.
\sedit{The first term is the regularization term that prevents numerical inflation of encoder outputs and training collapse.
The second term}
guides $\bm{\Sigma}_\mathrm{post}$ to have a similar power to the target waveform $\bm{x}_0$.

During inference steps of the diffusion model, the model can sample the initial noise from the prior $\mathcal{N}(0, \bm{\Sigma}_\mathrm{prior})$. Since the prior contains information related to the target waveform, it is expected to facilitate waveform generation compared to the naive approach of sampling from Gaussian noise.

In experiments~\cite{restoregrad}, RestoreGrad showed superior results in speech enhancement and image restoration compared to the hand-crafted method~\cite{priorgrad} and sampling from standard normal distribution.

\section{Proposed method}
\subsection{Motivation} \label{sec:motivation}
We hypothesize that the noise sampling issues of WaveFit described in Sec.~\ref{sec:ssl-wavefit} \redit{can} be addressed by introducing the trainable priors in RestoreGrad described in Sec.~\ref{sec:restoregrad}. In addition, imposing a new constraint on the trainable prior to match the energy of the target speech enables reference-aware gain adjustment, thereby resolving the gain adjustment issue.
These improvements are expected to reduce the difficulty of waveform modeling from data-driven features, enabling high-quality waveform generation with fewer inference steps.

\subsection{Model overview}
\begin{figure}[t!]
\begin{center}
\includegraphics[width=\linewidth]{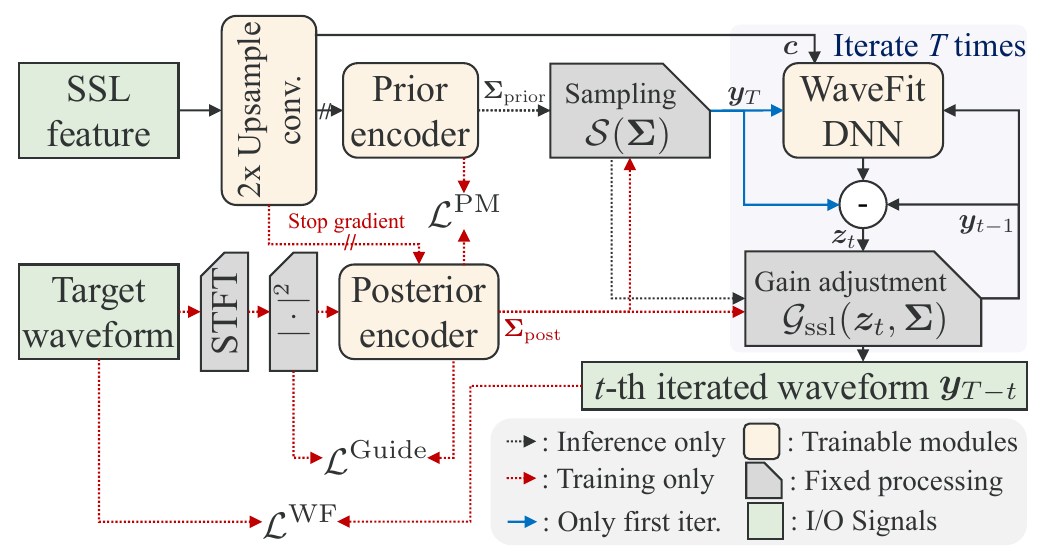}
\vspace{-8mm}
\caption{
Overview of the proposed model.
During training, the posterior encoder derived from the target waveform and the SSL feature is used for noise sampling and gain adjustment.
During inference, the prior encoder derived from the SSL feature is used for same process.
Solid arrows are used for both training and inference.
} \label{fig:model_overview}
\vspace{-9mm}
\end{center}
\end{figure}
Figure \ref{fig:model_overview} shows an overview of the proposed method. 
The proposed method introduces a prior encoder and a posterior encoder to achieve a trainable initial noise sampling \sssedit{as in RestoreGrad.
For the conditional input $c$, we used SSL features up-sampled by a factor of two using a transposed 2D convolutional layer as in \cite{miipher,miipher2}. $\bm{c}$ is used as the input of the posterior encoder, the prior encoder and the WaveFit DNN.}
\sssedit{During training, initial noise is sampled from the posterior distribution  $\mathcal{N}(0, \bm{\Sigma}_\mathrm{post})$, which is conditioned on the SSL feature and the target waveform. During inference, it is sampled from the prior distribution  $\mathcal{N}(0, \bm{\Sigma}_\mathrm{prior})$, which is conditioned solely on SSL features.}

\subsection{Noise sampling \sssedit{in time-frequency domain}} \label{sec:noise_sampling}
While RestoreGrad introduced trainable priors in the waveform domain, the proposed method incorporates them in the time-frequency domain to shorten the sequence length and reduce the modeling complexity.
Specifically, the variance feature $\bm{\Sigma}$ obtained from either the posterior or the prior encoder, is changed to have a shape of $\mathbb{R}^{F \times K}$. \sssedit{Here, $F$ is the frequency bin size and $K$ is the number of frames}. Then, the time domain initial noise $\bm{y}_{T}=\mathcal{S}(\bm{\Sigma}) \in \mathbb{R}^{D}$ is sampled with Eq.~(\ref{eq:proposed_noise_sampling}): 
\begin{align}
    \mathcal{S}(\bm{\Sigma})&=\mathrm{iSTFT}(\mathcal{R}(\bm{N})\odot \bm{\Sigma} + i \mathcal{I}(\bm{N})\odot\bm{\Sigma}), \label{eq:proposed_noise_sampling}\\
    \bm{N}&=\mathrm{STFT}(\epsilon) \in \mathbb{C}^{F\times K},~ 
    \epsilon \in \mathbb{R}^D \sim \mathcal{N}(0,\bm{I}).
\end{align}
This process corresponds to the reparameterization trick in a complex-valued VAE~\cite{cvae} when assuming zero mean and zero pseudo-covariance.
$\mathcal{R}(\cdot)$ and $\mathcal{I}(\cdot)$ are operators that extract only the real and imaginary parts, respectively.

\subsection{Loss function and gain adjustment} \label{sec:loss_func}
The loss function of the proposed method is as follows:
\begin{align}
    \mathcal{L}^\mathrm{TrainerFit} &= 
    \mathcal{L}^\mathrm{WF} 
    + \lambda_\mathrm{PM} \mathcal{L}^\mathrm{PM}
    + \mathcal{L}^\mathrm{Guide}.
\end{align}
Here, \sedit{the first term} is the same loss function as in Eq~(\ref{eq:wavefit_loss}), \sssedit{which is responsible for the training of diffusion and GAN.  
The second term is based on the loss term in Eq.~(\ref{eq:L_PM_restoregrad}), but is expanded to the time-frequency domain. This loss minimizes the KL divergence between the output of the prior encoder and that of the posterior encoder. $\lambda_\mathrm{PM}$ is the weight parameter.}
\sedit{Finally, the third term} softly provides guidance to the posterior encoder output $\bm{\Sigma}_\mathrm{post}$:
\begin{align}
    \mathcal{L}^\mathrm{Guide}\!\!&=\!
    \left|\mathcal{E}(\bm{\Sigma}_\mathrm{post})\!-\!\mathcal{E}(|\bm{X}_0|^2)\right| \!+\! \frac{\lambda_\mathrm{Guide}}{FK}\!\sum_{f=0}^{F-1}\sum_{k=0}^{K-1}\!\frac{\bm{\Sigma}_\mathrm{post}[f,k]}{|\bm{X}_0|^2[f,k]}. \label{eq:l_guide}
\end{align}
Here, $\lambda_\mathrm{Guide}$ is the weight parameter, $|\bm{X}_0|^2\in \mathbb{R}^{F\times K}$ is the power spectrogram of target waveform, and $f,k$ represent the \sssedit{frequency and time index, and \sssedit{$\mathcal{E}(\cdot)$ is a element-wise summation operation,} respectively.}
The first term aims to match the energy of the posterior encoder output with that of the target speech. This encourages $\bm{\Sigma}_\mathrm{post}$ to have the energy close to that of the target waveform. As a result, we can define a reference-aware gain adjustment operator as follows:
\begin{align}
    \mathcal{G}_\mathrm{ssl}(\bm{z}_t,\bm{\Sigma}) &= \sqrt{(\mathcal{E}(\bm{\Sigma})/(\mathcal{E}(|\bm{Z}_t|^2)+s))}\bm{z}_t.
\end{align}
Here, $s$ is a scalar to avoid zero-division.
The second term is an extension of the second term in Eq.~(\ref{eq:L_LR_restoregrad}) to two-dimensional signals, guiding effective posterior learning by softly reflecting the power of the target spectrogram.

\section{Experimental evaluation}
\begin{table*}[t]
\centering
\caption{
Evaluation results when using LibriTTS-R test-clean, 8-th layer SSL features, and $T=5$. 
\textbf{Bold} indicates the best method under the same conditions, and \underline{underlines} indicate significant differences between WaveFit and WaveTrainerFit.}
\label{table:overall_result}
\scalebox{0.98}[0.98]{
\begin{tabular}{cc|cccccc}
\hline \hline
\multicolumn{2}{c|}{Methods} & \multicolumn{4}{c}{Objective metrics} & \multicolumn{2}{c}{Subjective metrics} \\ 
SSL features & models & \begin{tabular}[c]{@{}c@{}}Speech BERT\\ Score (↑, \%)\end{tabular} & MCD (↓) & Log-F0-RMSE (↓) & \begin{tabular}[c]{@{}c@{}}Speaker\\ Similarity (↑, \%)\end{tabular} & N-MOS (↑) & S-MOS (↑) \\ \hline
\multicolumn{2}{c|}{Clean speech}
&  ($100.0$) 
&  ($0.0$) 
&  ($0.0$) 
&  ($100.0$) 
&  $3.92\pm0.10$ 
&  $-$ 
\\ \hline
\multirow{3}{*}{WavLM} 
& HiFi-GAN $V1$
&  $90.71$ 
&  $4.510$ 
&  $0.1972$ 
&  $49.09$ 
&  $2.39\pm0.12$ 
&  $2.81\pm0.12$ 
\\
& WaveFit
&  $94.28$ 
&  $4.109$ 
&  $0.1956$ 
&  $54.67$ 
&  $\bf\underline{3.76\pm0.11}$ 
&  $3.02\pm0.12$ 
\\
& WaveTrainerFit
&  $\bf95.28$ 
&  $\bf3.672$ 
&  $\bf0.1810$ 
&  $\bf62.61$ 
&  $3.50\pm0.11$ 
&  $\bf\underline{3.38\pm0.11}$ 
\\ \hline
\multirow{3}{*}{XLS-R} 
& HiFi-GAN $V1$
&  $91.09$ 
&  $4.424$ 
&  $0.1887$ 
&  $51.96$ 
&  $2.54\pm 0.12$ 
&  $2.99\pm 0.12$ 
\\
& WaveFit
&  $94.11$ 
&  $4.196$ 
&  $0.1934$ 
&  $52.78$ 
&  $\bf\underline{3.79\pm 0.11}$ 
&  $3.04\pm 0.12$ 
\\
& WaveTrainerFit 
&  $\bf94.39$ 
&  $\bf4.089$ 
&  $\bf0.1762$ 
&  $\bf55.54$ 
&  $3.21\pm 0.12$ 
&  $\bf3.13\pm 0.12$ 
\\ \hline
\multirow{3}{*}{\begin{tabular}[c]{@{}c@{}}Whisper\\ encoder\end{tabular}} 
& HiFi-GAN $V1$
&  $88.90$ 
&  $4.446$ 
&  $0.1843$ 
&  $54.98$ 
&  $2.41\pm 0.12$ 
&  $2.96\pm 0.12$ 
\\
& WaveFit
&  $93.30$ 
&  $3.715$ 
&  $0.1695$ 
&  $59.64$ 
&  $3.23\pm 0.12$ 
&  $3.56\pm 0.11$ 
\\
& WaveTrainerFit
&  $\bf94.60$ 
&  $\bf3.208$ 
&  $\bf0.1690$ 
&  $\bf75.02$ 
&  $\bf\underline{3.87\pm 0.10}$ 
&  $\bf\underline{4.19\pm 0.09}$ 
\\ \hline \hline
\end{tabular}
}
\end{table*}
\begin{figure}[t!]
\begin{center}
    \includegraphics[width=\linewidth]{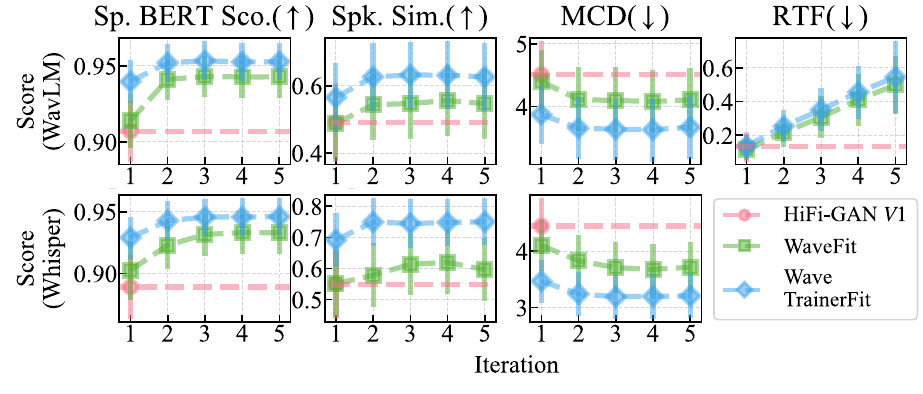}
    \vspace{-9mm}
  \caption{
  Line plots of objective metrics over iterations.
  \nsedit{Real-Time Factor (RTF)} was measured on ``Intel(R) Xeon(R) Silver 4316 CPU @ 2.30GHz'' using \nsedit{randomly selected} 200 samples.
  }
  \vspace{-5mm}
  \label{fig:intermediate_objective}
\end{center}
\end{figure}
\begin{table}[t]
\centering
\caption{Layer-focused objective evaluation results. Here, we used WavLM features as conditional features.} \label{tab:layer-wise_evaluation}
\scalebox{0.8}[0.8]{
\begin{tabular}{c|cccc|c}
\hline \hline
Models-Layers
& \begin{tabular}[c]{@{}c@{}}Sp. BERT\\ Score\end{tabular} 
& MCD
& \begin{tabular}[c]{@{}c@{}}Log-F0\\ RMSE\end{tabular} 
& \begin{tabular}[c]{@{}c@{}}Spk.\\ Sim.\end{tabular} 
& UTMOS ($\uparrow$)
\\ \hline
WaveFit-2
&  $95.95$ 
&  $\bf2.852$ 
&  $\bf0.1660$ 
&  $78.99$ 
&  $4.230$ 
\\
WaveTrainerFit-2
&  $\bf96.13$ 
&  $2.860$ 
&  $0.1662$ 
&  $\bf79.65$
&  $4.203$ 
\\ \hline
WaveFit-8
&  $94.28$ 
&  $4.109$ 
&  $0.1956$ 
&  $54.67$ 
&  $4.194$
\\
WaveTrainerFit-8
&  $\bf95.28$ 
&  $\bf3.672$ 
&  $\bf0.1810$ 
&  $\bf62.61$ 
&  $4.160$
\\ \hline
WaveFit-24
&  $89.08$ 
&  $5.501$ 
&  $0.3623$ 
&  $35.09$ 
&  $\underline{3.681}$
\\
WaveTrainerFit-24
&  $\bf92.19$ 
&  $\bf5.066$ 
&  $\bf0.2301$ 
&  $\bf40.76$ 
&  $\underline{\bf4.198}$
\\ \hline \hline
\end{tabular}
}
\vspace{-2mm}
\end{table}
\subsection{Experimental setup}
\noindent\textbf{Datasets.}
We used the LibriTTS-R corpus~\cite{libritts_r}. 
This corpus consists of a total of $585$ hours of speech at $24$kHz from $2,456$ speakers and text pairs.
The ``train-clean-360'', ``dev-clean'', and ``test-clean'' were used for training, validation, and test set, respectively.
\\\textbf{Model and training setup.}
\sedit{Our model generates waveforms by upsampling SSL features by $480 \times$.
The breakdown is as follows: First, as shown in Fig.~\ref{fig:model_overview}, a 2D transposed convolutional layer perform $2\times$ upsampling. 
Then, an architecture similar to WaveFit~\cite{wavefit} with \kedit{upsampling scale} of $\{5,4,3,2,2\}$ performs $240\times$.}
The model was trained with a batch size of $8$ for $400k$ training steps.
The weight parameters $\lambda_\mathrm{Guide}$ and $\lambda_\mathrm{PM}$ were set to $0.1$ and $10$, respectively.
For the encoder architecture, real-valued DCUnet-10~\cite{dcunet} was adopted with two primary modifications. 
First, \sedit{a linear layer was used to align the dimension of SSL features with the frequency dimension of the power spectrogram.} 
Second, two U-Nets~\cite{dcunet} were defined in the posterior encoder: one used for the power spectrogram from the target waveform, and the other for the SSL features.
The intermediate outputs of the first U-Net at each block were added to those of the second U-Net at the same position.
\kedit{The channel size of the encoders} was $32$ for the posterior encoder and $45$ for the prior encoder.
\kedit{The total number of parameters was $2.61~M$ and $2.59~M$, respectively.}
Other settings\kedit{,} such as optimizer, scheduler, and discriminator are the same as in the open-sourced implementation~\footnote{\scriptsize\url{https://github.com/yukara-ikemiya/wavefit-pytorch}}.
\\\textbf{Comparison methods.}
\sedit{To evaluate the effectiveness of the proposed method, we set up two baselines. 
In addition to WaveFit~\cite{miipher}, HiFi-GAN~\cite{hifigan} was set as a baseline to clarify the effectiveness of the WaveFit architecture itself in waveform generation from SSL features.
To align the \kedit{upsamping scale} to $240$, the upsample-rates were changed to $\{8,5,3,2\}$, and the upsample-kernel-sizes were changed to $\{16,11,7,4\}$. 
The total number of parameters for the generator was $17.18~M$.
The model was trained for $1M$ steps with the same batch size as the proposed method.
}
\\\textbf{Evaluation metrics.}
We adopted the following objective metrics: 
\sedit{SpeechBERTScore~\cite{sp_bert_score} is an automatic speech evaluation metric that measures semantic congruence between the generated speech and the reference speech from the BERTScore~\cite{bertscore} of SSL features.}
\kedit{Mel-cepstrum distortion (MCD) was computed to assess the spectral distortion.}
\kedit{Log-F0-RMSE was calculated by~\cite{harvest} to evaluate pitch accuracy.}
The above three metrics were calculated using the open-sourced implementation~\footnote{\scriptsize\url{https://github.com/Takaaki-Saeki/DiscreteSpeechMetrics}}.
Finally, speaker similarity is measured as the cosine similarity of speaker embeddings~\cite{ecapa_tdnn,speechbrain} between the target and the predicted waveform.
\sedit{To evaluate the quality of the generated waveforms, we conducted two subjective listening tests. 
A total of 450 samples per method were evaluated on a five-point scale by 15 participants.
To eliminate the effects of differences in waveform energy in all evaluations, all waveforms were aligned to the energy of clean speech.}
\\\textbf{Conditional features.}
\sedit{We used two SSL features and one ASR-driven feature as the conditional features of the vocoders: }
1. \textbf{WavLM-large}~\cite{wavlm} is an English-specialized SSL model pre-trained with masked language modeling objective. 
2. \textbf{XLS-R-0.3B}~\cite{xls_r} is a multilingual SSL model pre-trained with contrastive objective. 
3. \textbf{Whisper-medium}~\cite{whisper} is a multilingual speech-to-text model. 
Here, we used the encoder part of Whisper as a feature extractor similar to~\cite{whisper_speech_assessment,whisper_utt_retrieval}. 
\sedit{These features all come from different training objectives and pre-trained data, and are expected to have different properties.}

\subsection{Experimental results}
\vspace{-1mm}
\subsubsection{\kedit{Objective and subjective evaluation results}} \vspace{-1mm}
Table~\ref{table:overall_result} shows the evaluation results.
It can be seen that the proposed method outperformed the baseline on all reference-aware objective metrics regardless of SSL features.
\sssedit{This suggests that the proposed initial noise sampling and gain adjustment with trainable priors effectively worked to improve the objective metrics.}
Regarding the S-MOS, the proposed method again outperformed baselines for all the SSL features.
The reasons are discussed as follows: the spectral characteristics of the target speech is learned through the second term of Eq.~(\ref{eq:l_guide}) in the \kedit{prior} encoder.
As a result, initial noise containing pitch information can be sampled as shown in Fig.~\ref{fig:concept}~(b), which led to waveform generation that well captures speaker characteristics.
For the N-MOS, the proposed method outperformed the baselines with the Whisper encoder, but fell short when using XLS-R.
There are several possible causes.
\sedit{One possibility is that the proposed method is sensitive to hyperparameters.
RestoreGrad, which our method is based on, showed a tendency to be somewhat sensitive to the weight parameter of $\mathcal{L}^\mathrm{LR}$~\cite{restoregrad}.
Due to computational cost, we tested only the weight that achieved the highest score in RestoreGrad, but careful tuning of these weights may yield better results.}
\sedit{Another possibility is that features of XLS-R themselves had a negative impact.
XLS-R shows worse performance compared to the baseline in English ASR tasks, suggesting language interference within the model~\cite{xls_r}.
A similar phenomenon may have occurred in our experiments, possibly leading to degraded performance.}
An important future work is a detailed analysis of how combinations of these \sedit{hyperparameters}, languages, SSL features, and other factors affect performance.
\subsubsection{Performance at each iteration and processing speed} \vspace{-1mm}
\redit{Figure~\ref{fig:intermediate_objective} shows the objective metrics for each inference iteration.} 
\kedit{During inference, the real-time factor (RTF) of the proposed method was slightly increased by the prior encoder and the sampling process; nevertheless, it showed superior scores in all metrics except for RTF across all iterations.}
Specifically, the SpeechBERTScore at iteration 1 shows the largest improvement margin for the proposed method.
These results suggest that inference starting from initial noise close to the target waveform leads to better waveform generation compared to the baseline with the same number of iterations. 

\vspace{-1.5mm}
\subsubsection{Performance for \kedit{SSL} features from different layers} \vspace{-1mm}
Features from shallow layers \kedit{of SSL models} are known to contain much acoustic information from the input, while those from deeper layers contain much semantic information~\cite{ssl_layer_wise_analysis}.
To verify the robustness of the proposed method when conditioning on features with different properties, we conducted evaluations using WavLM features extracted from the 2nd, 8th, and 24th layers.
In this experiment, reference-free UTMOS~\cite{utmos} was added as an evaluation metric that excludes speaker characteristic deviations from the target waveform.

The results are shown in Table~\ref{tab:layer-wise_evaluation}.
It can be seen that the proposed method shows better results than the baseline in all reference-aware metrics except for two metrics in the 2nd layer. 
\kedit{
Notably, when conditioned on features from the 24th layer, which contain limited acoustic information, the proposed method achieved substantial improvements in reference-aware metrics in addition to the aforementioned advantages, without significant degradation in UTMOS.
This suggests that the proposed method is capable of generating waveforms more consistent with the reference speech, thereby exhibiting stronger robustness to variations in conditional features.
}
\vspace{-2mm}
\section{Conclusion}
We proposed WaveTrainerFit for high-quality waveform generation from SSL features. 
By introducing a VAE-based trainable prior, we achieved superior initial noise sampling and reference-based gain adjustment, addressing the problems of conventional WaveFit from SSL features. 
In evaluation experiments, we showed that the proposed method can generate high-quality waveforms from features lacking acoustic information and with fewer iterations. 
In subjective evaluations, we confirmed that the proposed method consistently showed better results in speaker similarity MOS compared to the baseline. 
Future work includes building models from large datasets and extending to multilingual models.

\vfill\pagebreak
\bibliographystyle{IEEEtran}
\bibliography{refs}

\end{document}